\documentclass[12pt]{iopart}
\usepackage{graphicx}
\usepackage{bm}

\begin{document}

\title[Hole spin relaxation in QDs]{Spin-orbit induced hole spin relaxation in InAs and GaAs quantum dots}

\author{J I Climente, C Segarra and J Planelles}

\address{Departament de Qu\'{\i}mica F\'{\i}sica i Anal\'{\i}tica, Universitat Jaume I, Box 224, E-12080 Castell\'o, Spain}

\ead{josep.planelles@qfa.uji.es}

\date{\today}

\begin{abstract}
We study the effect of valence band spin-orbit interactions on the acoustic phonon assisted
spin relaxation of holes confined in quantum dots. Heavy hole-light hole (\emph{hh-lh}) mixing and all 
the spin-orbit terms arising from zinc-blende bulk inversion asymmetry (BIA) are considered on equal 
footing in a fully 3D Hamiltonian. We show that \emph{hh-lh} mixing and BIA have comparable contributions to
the hole spin relaxation in self-assembled QDs, but BIA becomes dominant in gated QDs.
Simultaneously accounting for both mechanisms is necessary for quantitatively correct results in quasi-2D QDs.
The dependence of the hole spin relaxation on the QD geometry and spin splitting energy is
drastically different from that of electrons, with a non-monotonic behavior which results from 
the interplay between SOI terms. Our results reconcile contradictory predictions of previous theoretical works 
and are consistent with experiments.
\end{abstract}

\submitto{\NJP}
\maketitle

\section{Introduction}

In the last years, the spin of holes confined in III-V semiconductor quantum dots (QDs) 
has emerged as a promising building block for spintronic and spin-based quantum information 
devices.\cite{FischerSSC} As compared to electrons, the $p$-like nature of the hole 
orbitals leads to weaker hyperfine interaction with the lattice nuclei, resulting
in coherence times which hold promise for applications.\cite{GerardotNAT,BrunnerSCI,FallahiPRL,ChekhovichPRL,EblePRL,TestelinPRB}
As a matter of fact, demonstrations of hole spin manipulation in QDs have been recently 
reported\cite{DeGreveNP,GoddenPRL,GreilichNP} and theoretical proposals of 
control mechanisms are being proposed.\cite{SzumniakPRL,HsiehPRB,BudichPRB,RoloffNJP}
In this context, the study of hole spin relaxation has become a subject of interest.
Hole spin relaxation is also important for optical applications because it is believed
to rule the exciton spin dynamics in both dark-to-bright exciton transitions\cite{LiaoPRB,KurtzePRB,KowalikPRB,HuxterCPL} 
--which affect exciton storage times\cite{LundstromSCI,ReischlePRL,LinOE}--
and transitions within the bright doublet\cite{TsitsishviliPRB10} --which 
affect light depolarization\cite{KowalikPRB}--.

Experimental observations in self-assembled InAs QDs point at hole spin lifetimes
ranging from $T_1^h \sim 10$ ps to $T_1^h \sim 1$ ms.\cite{GerardotNAT,GundogduAPL,HallAPL,LaurentPRL,HeissPRB}
The large dispersion is partly attributed to the different relaxation mechanisms involved in 
different studies. When the energy splitting between orthogonal spin states is small, 
hyperfine interaction is the dominant relaxation channel.\cite{KurtzePRB,FrasPRB}
In this case, the lifetime is strongly dependent on the degree of \emph{hh-lh} mixing. 
If the hole state is a pure \emph{hh}, as in the ground state of flat (quasi-2D) QDs, 
the hyperfine interaction takes an Ising form and spin relaxation is slow, but it rapidly 
increases in non-flat QDs due to \emph{hh-lh} mixing.\cite{FischerSSC,TestelinPRB}
On the other hand, when the energy splitting exceeds the nuclear magnetic field, the valence band
spin-orbit interaction (SOI) takes over as the main source of relaxation.\cite{KurtzePRB,FrasPRB}
Long hole spin lifetimes have then been observed, reaching up to $T_1^h \sim 0.25$ ms,
which is only 5-10 times shorter than electron spin lifetimes, $T_1^e$.\cite{HeissPRB}
This result is encouraging for the use of holes in quantum information and optical
applications, but it is surprising because the valence band SOI interaction is known 
to be much stronger than that of the conduction band.\cite{Winkler_book}

The above paradox has prompted a number of theoretical works trying to understand
which factors determine the relaxation dynamics of single holes under magnetic 
fields\cite{WoodsPRB,LuPRB,BulaevPRL,TrifPRL,WeiPRB} and that of holes forming 
excitons\cite{LiaoPRB,TsitsishviliPRB10,RoszakPRB,TsitsishviliPRB} in quasi-2D 
InAs/GaAs QDs. For the relaxation to take place one needs a source of 
energy relaxation, which in these systems is provided by the acoustic phonon 
bath,\cite{GundogduAPL,HallAPL} plus a source of spin admixture. 
Woods et al.\cite{WoodsPRB} and Lu et al.\cite{LuPRB} proposed that the latter
is the coupling between \emph{hh} and \emph{lh} subbands. Other authors have suggested 
instead that the splitting between \emph{hh} and \emph{lh} subbands in flat QDs is large 
owing to confinement and strain, so that spin admixture must be due to other SOI mechanisms. 
It was then proposed that hole SOI should have a similar origin to that of conduction electrons, 
namely the BIA of zinc-blende crystals, which gives rise to Dresselhaus SOI terms.\cite{Winkler_book} 
Bulaev and Loss assumed the cubic-in-$k$ Dresselhaus term is dominant and showed that 
$T_1^h$ could then become comparable to $T_1^e$ in flat QDs.\cite{BulaevPRL}
Other studies followed this assumption and succeeded in explaining some experimental observations.\cite{HeissPRB,TrifPRL,LiaoPRB} 
By contrast, Tsitsishvili et al. suggested that if the lateral confinement is weak,
it is the linear-in-$k$ term that dominates the mixing.\cite{TsitsishviliPRB} 
Last, Roszak et al.\cite{RoszakPRB} suggested that for holes forming excitons
it is the electron-hole (\emph{e-h}) exchange interaction together with the strain that gives
rise to hole spin admixture.

It is worth noting that all the previous works assumed a dominating SOI 
mechanism without actually comparing with others. In addition, simplified models
disregarding \emph{hh-lh} mixing become highly parametric, and different 
parameters were needed to explain different experimental observations even in 
the same system.\cite{HeissPRB,TrifPRL,LiaoPRB}
The lack of a comprehensive study translates into many open questions which
show that hole spin relaxation in QDs is still not fully understood.
To name a few: 
(i) while Ref.~\cite{WoodsPRB} predicts that $T_1^h$
decreases with the QD diameter, Ref.~\cite{LuPRB} 
predicts exactly the opposite behavior; 
(ii) Ref.~\cite{BulaevPRL} predicted $T_1^h > T_1^e$ in the limit of 2D QDs, 
but experiments on self-assembled QDs have only shown $T_1^e/T_1^h=5-10$,\cite{HeissPRB} 
so that one wonders if any realistic QD structure would actually permit holes relaxing slower than electrons;
(iii) in excitons, the role of \emph{e-h} exchange energy is not clear: 
while experiments with self-assembled QDs have shown neglegible dependence of $T_1^h$,\cite{HallAPL} 
a strong dependence has been found in colloidal quantum rods.\cite{HePRL}

In this work, we aim at covering the existing gap in the understanding of
hole spin relaxation in QDs. We study the hole spin dynamics considering
simultaneously the most relevant intrinsic SOI terms of III-V QDs,
namely \emph{hh-lh} mixing and all the different Dresselhaus SOI terms arising 
from the BIA of zinc-blende crystals, along with the hole-acoustic phonon
coupling.  All the terms are described within a four-band k$\cdot$p formalism 
and 3D Hamiltonians, which allows us to provide a general overview on the effect
of the QD size and geometry dependence while relying on well-known bulk parameters only.
In this way, we are able to establish the regime of validity of previous
studies which assumed a single dominating SOI mechanism. Furthermore, we
explore spheroidal QDs beyond the usual quasi-2D limit, thus providing 
theoretical assessment for developing experimental research with spherical and 
prolate QDs.\cite{HuxterCPL,HePRL}

We find that \emph{hh-lh} is the main SOI channel in prolated or spherical QDs,
but Dresselhaus SOI has a comparable contribution in oblated QDs (such as 
self-assembled dots), and it becomes dominant in quasi-2D QDs with very weak 
confinement (as in electrostatically confined dots). 
The competition between SOI coupling terms and the energy splitting between
\emph{hh} and \emph{lh} leads to a non-monotonic dependence of $T_1^h$ with the QD geometry,
in sharp contrast with the well-known case of electrons.
This explains the opposite trends reported by different theoretical studies in the literature.
The dependence of $T_1^h$ on the electron-hole exchange energy we predict is consistent 
with experiments on colloidal nanorods\cite{HePRL}, but it suggests that 
two-phonon processes are relevant in self-assembled QDs.
In prolate QDs, where the ground state is formed by \emph{lh}, the spin
relaxation is shown to take place in similar timescales as for transitions
between \emph{hh} states. However, the coupling to acoustic phonons is 
different, with deformation potential interaction being the main mechanism even 
for vanishing spin splitting energy.

\section{Theoretical Model}

We study the spin relaxation of holes confined in zinc blende QDs grown along
the [001] direction. 
The hole spin states are considered split energetically, for example by the 
\emph{e-h} exchange interaction in excitons or any other source that can be 
viewed as an effective axial magnetic field. Thus, similar results can be expected 
for transitions between Zeeman sublevels under moderate external magnetic fields.
\begin{figure}[h]
\begin{center}
\includegraphics[width=0.45\textwidth]{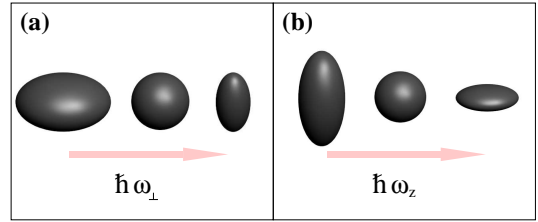}
\caption{Geometry of QDs with varying lateral (a) and vertical (b) confinement frequency.}
\label{fig:intro}
\end{center}
\end{figure}
\subsection{Hamiltonian}
The system Hamiltonian reads:
\begin{equation}
H \,=\, H_h + H_{ph} + H_{h-ph}.
\label{eq:H}
\end{equation}
\noindent In Eq.~(\ref{eq:H}), $H_h$ is the hole Hamiltonian,
\begin{equation}
H_h \,=\, H_{L} + H_{BIA} + V_{QD}\,{\cal I} + H_Z, 
\label{eq:Hh}
\end{equation}
\noindent where $H_{L}$ is the 4-band Luttinger Hamiltonian describing
the coupled \emph{hh-lh} bands.\cite{Luttinger}  It includes quadratic
terms in $k$ only:
\begin{eqnarray}
\label{lutt1}
\fl H_L  =\frac{1}{m_0} [(\gamma_1+\frac{5}{2} \gamma_2) \frac{k^2}{2}  
       - \gamma_2 (k_x^2 J_x^2+k_y^2 J_y^2+k_z^2 J_z^2)  \nonumber \\
	 -2 \gamma_3 (\{k_x,k_y\} \{J_x,J_y\}+\{k_y,k_z\} \{J_y,J_z\} %\nonumber \\
	 +\{k_z,k_x\} \{J_z,J_x\})]
\end{eqnarray}

\noindent where $m_0$ is the free electron mass, $\gamma_i$ are the Luttinger parameters,
$k_j=-i\,\hbar\,\partial_j$ the $j$ component of the linear momentum,
$\{A,B\}=\frac{1}{2}(AB+BA)$ and $J_i$ is the i-th component of the angular momentum 
corresponding to the quantum number $J=3/2$. To get the matrix representation of this hamiltonian
we multiply the first term of equation (\ref{lutt1}) by the 4 x 4 unit matrix and employ the 
standard matrix representation of the $J=3/2$ components of the angular momentum.\cite{Voon_book} 
We finally get:
\begin{equation}
H_L \,=\,
\left( 
\begin{array}{cccc} 
P + Q        &  -S         &   R         &  0 \\
-S^{\dagger} & P-Q         &   0         &  R \\
R^{\dagger}  &   0         &  P-Q        &  S \\
0            & R^{\dagger} & S^{\dagger} & P+Q
\end{array}
\right),
\label{eq:Hl}
\end{equation}
\noindent with
\begin{eqnarray}
\label{eq:P}
P &=& \frac{1}{2\,m_0}\,\gamma_1\,\left( k_x^2 + k_y^2 + k_z^2 \right), \\
\label{eq:Q}
Q &=& \frac{1}{2\,m_0}\,\gamma_2\,\left( k_x^2 + k_y^2 - 2 k_z^2 \right),  \\
\label{eq:R}
R &=& \frac{1}{2\,m_0}\,\left[ -\sqrt{3}\,\gamma_2\,(k_x^2-k_y^2) + 2\,i\,\sqrt{3}\,\gamma_3\,k_x\,k_y \right]  \\
\label{eq:S}
S &=& \frac{1}{2\,m_0}\,2\,\sqrt{3}\,\gamma_3\,(k_x -i\,k_y)\,k_z.
\end{eqnarray}
$H_{BIA}$ includes the linear and the Dresselhaus SOI third order in $k$ terms:\cite{Winkler_book}
\begin{eqnarray}\label{eq:Hbia}
H_{BIA}  &=  \frac{2}{\sqrt{3}}\,C_k\,[ k_x\,\{J_x,J_y^2-J_z^2\} + \mbox{cp} ]  \nonumber \\
         &+  b_{41}\,(\{k_x,k_y^2-k_z^2\}\,J_x + \mbox{cp})  \nonumber \\
         &+  b_{42}\,(\{k_x,k_y^2-k_z^2\}\,J_x^3 + \mbox{cp})  \nonumber \\
         &+  b_{51}\,(\{k_x,k_y^2+k_z^2\}\,\{J_x,J_y^2-J_z^2\}+\mbox{cp})  \nonumber \\
         &+  b_{52}\,(k_x^3\,\{J_x,J_y^2-J_z^2\}+\mbox{cp}),
\end{eqnarray}

\noindent where $C_k$, $b_{41}$, $b_{42}$, $b_{51}$ and $b_{52}$ are material dependent
coefficients and $\mbox{cp}$ stands for cyclic permutations of the preceding terms. 
 The matrix form of the Hamiltonian terms above is given in the Appendix. One can notice that 
 all BIA terms provide direct mixing between \emph{hh} spin up and spin down ($J_z=+3/2$ and $J_z=-3/2$) 
components except for $b_{41}$, which requires the participation of the \emph{lh} 
($J_z=+1/2$ and $J_z=-1/2$) components.  
Rashba SOI is neglected in this study because it is an extrinsic effect, which depends on 
the details of the electric field felt by the system. Besides, for holes it couples energetically 
distant states so that, under moderate external fields, it is less efficient than Dresselhaus SOI.\cite{BulaevPRL}
$V_{QD}$ describes the confining potential of the QD.
We model QDs with parabolic confinement in the $x,\,y$ and $z$ directions,
\begin{equation}
V_{QD} \,=\, -\frac{1}{2}\,\chi_{\perp}\,(x^2+y^2) - \frac{1}{2}\,\chi_{z} z^2
\label{eq:pot}
\end{equation}
\noindent where $\chi_{\perp}$ and $\chi_{z}$ are the force constants 
perpendicular and parallel to the growth direction, respectively. 
Eq.~(\ref{eq:pot}) allows us to simulate 3D spheroidal QDs with different aspect 
ratios, from flat (quasi-2D) to spherical or elongated (quasi-1D) structures, see 
Fig.~\ref{fig:intro}.
$H_Z$ is the Hamiltonian modeling the splitting of the hole states by an effective axial 
magnetic field, three times larger for heavy ($|J_z|=3/2$) than for light holes ($|J_z|=1/2$).
This field could originate e.g. from the \emph{e-h} exchange interaction\cite{RoszakPRB}
or a spin Zeeman effect. Then, we assume that:
\begin{equation}
H_Z \,=\, \frac{1}{2}\,
\left( 
\begin{array}{cccc} 
\Delta        &   0         &   0         &  0 \\
0            & \frac{1}{3}\,\Delta         &   0         &  0 \\
0            &   0         &  -\frac{1}{3}\,\Delta        &  0 \\
0            &   0         &   0           & -\Delta
\end{array}
\right).
\end{equation}
To calculate the hole states from $H_h$, we note that the diagonal terms
correspond to harmonic oscillator Hamiltonians: % for \emph{hh} and \emph{lh}:
\begin{equation}
%P + Q + V_{QD} &=& \frac{\hbar^2}{2\,m_\perp^{hh}}\,k_{\perp}^2 - \frac{1}{2}\,\chi_\perp\,(x^2+y^2) \\
%      &+& \frac{\hbar^2}{2\,m_z^{hh}}\,k_{z}^2 - \frac{1}{2}\,\chi_z\,z^2, 
%P + Q + V_{QD} &=& T_{hh,\perp} - \frac{1}{2}\,\chi_\perp\,(x^2+y^2) %\\
%      &+& T_{hh,z} - \frac{1}{2}\,\chi_z\,z^2, 
P + Q + V_{QD} = T_{hh,\perp} - \frac{1}{2}\,\chi_\perp\,(x^2+y^2) + T_{hh,z} - \frac{1}{2}\,\chi_z\,z^2, 
\label{eq:diagHH}
\end{equation}
\noindent and
\begin{equation}
%P - Q + V_{QD} &=& \frac{\hbar^2}{2\,m_\perp^{lh}}\,k_{\perp}^2 - \frac{1}{2}\,\chi_\perp\,(x^2+y^2) \\
%      &+& \frac{\hbar^2}{2\,m_z^{lh}}\,k_{z}^2 - \frac{1}{2}\,\chi_z\,z^2, 
%P - Q + V_{QD} &=& T_{lh,\perp} - \frac{1}{2}\,\chi_\perp\,(x^2+y^2) \\
%      &+& T_{lh,z} - \frac{1}{2}\,\chi_z\,z^2, 
P - Q + V_{QD} = T_{lh,\perp} - \frac{1}{2}\,\chi_\perp\,(x^2+y^2) + T_{lh,z} - \frac{1}{2}\,\chi_z\,z^2, 
\label{eq:diagLH}
\end{equation}
\noindent where $T_{i,j}=\frac{\hbar^2}{2\,m_j^i}\,k_j^2$, with $i=(hh,\,lh)$, $j=(\perp,z)$,
$k_\perp=(k_x^2+k_y^2)^{1/2}$, $m_\perp^{hh}=m_0/(\gamma_1+\gamma_2)$, $m_z^{hh}=m_0/(\gamma_1-2\gamma_2)$, 
 $m_\perp^{lh}=m_0/(\gamma_1-\gamma_2)$, and $m_z^{lh}=m_0/(\gamma_1+2\gamma_2)$. % and $\omega_j^{i}=\sqrt{\chi_j/m_j^{i}}$.
This suggests rewriting all derivatives and coordinates of $H_h$ in terms of harmonic 
oscillator ladder operators and then projecting it onto a basis formed by oscillator 
eigenstates. %$|\nu_x,\,\nu_y,\,\nu_z\,\rangle$, where $\nu_j=0,1,2 \ldots$
The problem is that Eq.~(\ref{eq:diagHH}) has \emph{hh} masses while Eq.~(\ref{eq:diagLH}) has \emph{lh} masses,
and hence they have different oscillator frequency, $\omega_j^{i}=(\chi_j/m_j^{i})^{1/2}$.
Because the off-diagonal terms of $H_L$ couple \emph{hh} and \emph{lh} components, 
it is convenient to use a single kind of oscillator states, e.g. \emph{hh} states. 
This can be done by rewriting Eq.~(\ref{eq:diagLH}) in terms of the \emph{hh} harmonic
oscillator Hamiltonians: 

\begin{eqnarray}
\nonumber
P - Q + V_{QD} &=& \frac{\gamma_1-\gamma_2}{\gamma_1+\gamma_2}\,H_{hh,\perp} - \frac{\gamma_2}{\gamma_1+\gamma_2}\,\chi_\perp\,(x^2+y^2) \\ 
               &+& \frac{\gamma_1+2\gamma_2}{\gamma_1-2\gamma_2}\,H_{hh,z} + \frac{2\gamma_2}{\gamma_1-2\gamma_2}\,\chi_z\,z^2,
\end{eqnarray}

\noindent where $H_{hh,\perp}=T_{hh,\perp}-\chi_\perp\,(x^2+y^2)/2$ and 
$H_{hh,z}=T_{hh,z}-\chi_z\,(z^2)/2$.
The resulting hole states are Luttinger spinors of the form:
\begin{equation}\label{eq:psi}
|\Psi^h_m> =  \sum_{r,J_z} c_{r,J_z}^m | v_x, v_y, v_z \rangle\, |3/2,J_z \rangle,
\end{equation}
\noindent where $v_j=0,1,2\ldots$ is the quantum number of the 1D \emph{hh} harmonic oscillator along the
$j$ direction, $r$ is the combined orbital quantum number, $r=(v_x, v_y, v_z)$ and $|3/2,J_z\rangle$
the Bloch function. %$J_z=\pm 3/2$ corresponds to spin up and spin down \emph{hh} components, 
%while $J_z=\pm 1/2$ corresponds to \emph{lh} components.

%is the quantum number of the 1D harmonic oscillator along the $j$ direction.

$H_{ph}$ in Eq.~(\ref{eq:H}) is the Hamiltonian of acoustic phonons,
given by $H_{ph}=\sum_{\mathbf{q}\,\lambda}\,\hbar\,\omega_{q\,\lambda}\,a_{\mathbf{q}\,\lambda}^\dagger\,a_{\mathbf{q}\,\lambda}$, with
$\omega_{q\,\lambda}$ standing for the phonon energy spectrum of branch $\lambda$ 
($\lambda=l,t1,t2$ for longitudinal and the two transversal phonon modes)
and momentum $\mathbf{q}$.
We restrict to low phonon energies, where the linear dispersion regime applies,
$\omega_{q  \lambda}\,=\,c_{\lambda}\,q$, with $c_{\lambda}$
as the phonon velocity.

$H_{h-ph}$ is the hole-phonon interaction, 
\begin{equation}
H_{h-ph}=e\,\phi_{pz}\,{\cal I} + H_{dp},
\label{eq:h-ph}
\end{equation}
\noindent where $e$ is the hole charge, $\phi_{pz}$ the piezoelectric potential and $H_{dp}$ the 
deformation potential term. These are the two relevant scattering 
mechanisms at low temperatures.\cite{LuPRB} The piezoelectric potential is given by:\cite{UenoyamaPRB}
\begin{equation} \label{eq:pz}
\phi_{pz}  = \sum_{\lambda} \phi_{pz}^{\lambda} = 
 -\sum_{{\lambda}\,\mathbf{q}}\,\frac{4\pi\,i}{\epsilon_r\,q^2}\,h_{14}\,
\left( q_x\,\varepsilon_{yz}^\lambda + q_y\,\varepsilon_{zx}^{\lambda} + q_z\,\varepsilon_{xy}^{\lambda} \right).
\end{equation}
\noindent where $\epsilon_r$ is the relative dielectric constant, $h_{14}$ the piezoelectric 
constant and $\varepsilon_{ij}$ the strain tensor component.
On the other hand, the deformation potential term is given by the Bir-Pikus strain
Hamiltonian:
\begin{equation}\label{eq:dp}
{\cal H}_{dp} = \sum_\lambda \left( 
\begin{array}{cccc} 
p^\lambda + q^\lambda      &  -s^\lambda &   r^\lambda     &  0 \\
-(s^\lambda)^{\dagger} & p^\lambda-q^\lambda         &   0         &  r^\lambda \\
(r^\lambda)^{\dagger}  &   0         &  p^\lambda-q^\lambda        &  s \\
0            & (r^\lambda)^{\dagger} &  (s^\lambda)^{\dagger} & p^\lambda+q^\lambda
\end{array}
\right),
\end{equation}
\noindent where:
\begin{eqnarray}
p^\lambda &=& -a\,(\varepsilon_{xx}^\lambda + \varepsilon_{yy}^\lambda + \varepsilon_{zz}^\lambda), \\
q^\lambda &=& -\frac{b}{2}\,(\varepsilon_{xx}^\lambda + \varepsilon_{yy}^\lambda - 2\varepsilon_{zz}^\lambda), \\
r^\lambda &=& \frac{\sqrt{3}}{2}\,b\,(\varepsilon_{xx}^\lambda - \varepsilon_{yy}^\lambda) -i\,d\,\varepsilon_{xy}^\lambda, \\
s^\lambda &=& -d ( \varepsilon_{zx}^\lambda -i\,\varepsilon_{yz}^\lambda).
\end{eqnarray}
\noindent Here $a,\,b$ and $d$ are the valence band deformation potential
constants. 

The components of the strain tensor are rewritten using normal-modes coordinates,\cite{WoodsPRB}
\begin{equation}\label{eq:strain_q}
\varepsilon_{ij}^{\lambda}  = \, -\frac{i}{2}\,\sum_{\mathbf{q}} U^\lambda(q)\,
( \eta_\lambda^i(\mathbf{q})\,q_j +  \eta_\lambda^j(\mathbf{q})\,q_i )   %\\
% & \left(a_{\mathbf{q}}\,e^{i\,{\mathbf{q}}{\mathbf{r}}} + a_{\mathbf{q}}^{\dagger}\,e^{-i\,{\mathbf{q}}{\mathbf{r}}} \right).
  F(\mathbf{q},\mathbf{r}),
\end{equation}
\noindent where 
$F(\mathbf{q},\mathbf{r})=(e^{-i \mathbf{q} \mathbf{r}}\,a_q^+ + e^{i \mathbf{q} \mathbf{r}}\,a_q)$
and $U^\lambda(q)=\sqrt{{\hbar}/{2\,\rho\,V\,\omega_{q\,\lambda}}}$, with
$\rho$ and $V$ standing for the crystal density and volume. $\eta_\lambda(\mathbf{q})$
are the phonon polarization vectors:
$\mathbf{\eta}_l({\mathbf{q}}) = \left( q_x,\, q_y,\, q_z \right)/q$,
$\mathbf{\eta}_{t1}({\mathbf{q}}) = 
\left(q_x\,q_z, q_y\,q_z, -q_{\perp}^2 \right)/q\,q_{\perp}$ and
$\mathbf{\eta}_{t2}({\mathbf{q}}) = 
\left(  q_y, -q_x, 0 \right)/q_{\perp}$, with $q_{\perp}=\sqrt{q_x^2+q_y^2}$.
%At zero temperature the annihilation operator term in Eq.~(\ref{eq:strain_q}),
%associated with phonon absorption, is dropped.  
The piezoelectric potential now reads:
\begin{eqnarray}
\label{eq:ini}
\nonumber
\phi_{pz}^{l}\! &=&\! -\frac{12\,\pi\,h_{14}}{\epsilon_r}\,U^l(q)\,
\sum_{\mathbf{q}}\,\frac{q_x q_y q_z}{q^3}\,F(\mathbf{q},\mathbf{r}), \\
%\sum_{\mathbf{q}}\,\frac{q_x q_y q_z}{q^3}\,e^{-i \mathbf{q} \mathbf{r}}\,a_q^+,\\
%
\nonumber
\phi_{pz}^{t1}\!\! &=&\!\! -\frac{4\,\pi\,h_{14}}{\epsilon_r}\,U^t(q)\,
\sum_{\mathbf{q}}\,\frac{q_x q_y \,(2q_z^2 - q_{\perp}^2)}{q^3\,q_{\perp}}\,F(\mathbf{q},\mathbf{r}), \\
%\sum_{\mathbf{q}}\,\frac{q_x q_y \,(2q_z^2 - q_{\perp}^2)}{q^3\,q_{\perp}}\,e^{-i \mathbf{q} \mathbf{r}}\,a_q^+,\\
%
\phi_{pz}^{t2}\! &=&\! -\frac{4\,\pi\,h_{14}}{\epsilon_r}\,U^t(q)\,
\sum_{\mathbf{q}}\,\frac{q_z \,(q_y^2 - q_x^2)}{q^2\,q_{\perp}}\,F(\mathbf{q},\mathbf{r}),
%\sum_{\mathbf{q}}\,\frac{q_z \,(q_y^2 - q_x^2)}{q^2\,q_{\perp}}\,e^{-i \mathbf{q} \mathbf{r}}\,a_q^+.
\end{eqnarray}
\noindent %where $F(\mathbf{q},\mathbf{r})=(e^{-i \mathbf{q} \mathbf{r}}\,a_q^+ + e^{i \mathbf{q} \mathbf{r}}\,a_q)$.
In turn, the deformation potential operators become:
\begin{eqnarray}\label{eq:dpl}
\nonumber
p^l &=& i\,a\,U^l(q)\, \sum_{\mathbf{q}} 
\,q\,
\,F(\mathbf{q},\mathbf{r}), \\
%\,e^{-i\,{\mathbf{q}}{\mathbf{r}}}\, a_{\mathbf{q}}^\dagger,\\
\nonumber
q^l &=& i\,\frac{b}{2}\,U^l(q)\, \sum_{\mathbf{q}} 
\left( q - 3\,\frac{q_z^2}{q} \right)
\,F(\mathbf{q},\mathbf{r}), \\
%\,e^{-i\,{\mathbf{q}}{\mathbf{r}}}\, a_{\mathbf{q}}^\dagger,\\
\nonumber
r^l &=& -i\,U^l(q)\, \sum_{\mathbf{q}} 
\left( \frac{\sqrt{3}}{2}b\,\frac{q_x^2-q_y^2}{q} - id\,\frac{q_x\,q_y}{q} \right)
\,F(\mathbf{q},\mathbf{r}), \\
%\,e^{-i\,{\mathbf{q}}{\mathbf{r}}}\, a_{\mathbf{q}}^\dagger,\\
s^l &=& i\,d\,U^l(q)\, \sum_{\mathbf{q}} 
\frac{q_z\,(q_x-i\,q_y)}{q}
\,F(\mathbf{q},\mathbf{r}), 
%\,e^{-i\,{\mathbf{q}}{\mathbf{r}}}\, a_{\mathbf{q}}^\dagger,
\end{eqnarray}
\noindent for longitudinal phonons,
\begin{eqnarray}\label{eq:dpt1}
\nonumber
p^{t1} &=& 0, \\
\nonumber
q^{t1} &=& i\,\frac{b}{2}\,U^{t}(q)\, \sum_{\mathbf{q}} 
\left( \frac{3\,q_z\,q_{\perp}}{q} \right)
\,F(\mathbf{q},\mathbf{r}), \\
%\,e^{-i\,{\mathbf{q}}{\mathbf{r}}}\, a_{\mathbf{q}}^\dagger,\\
\nonumber
%r^{t1} &=& -i\,U^{t}(q)\, \sum_{\mathbf{q}} 
%\left( \frac{\sqrt{3}}{2}b\,\frac{q_z\,(q_x^2-q_y^2)}{q\,q_\perp} - id\,\frac{q_x\,q_y\,q_z}{q\,q_\perp} \right) \\
%\nonumber
% & & \,F(\mathbf{q},\mathbf{r}), \\
r^{t1} &=& -i\,U^{t}(q)\, \sum_{\mathbf{q}} 
\left( \frac{\sqrt{3}}{2}b\,\frac{q_z\,(q_x^2-q_y^2)}{q\,q_\perp} - id\,\frac{q_x\,q_y\,q_z}{q\,q_\perp} \right) \,F(\mathbf{q},\mathbf{r}), \\
%\,e^{-i\,{\mathbf{q}}{\mathbf{r}}}\, a_{\mathbf{q}}^\dagger,\\
s^{t1} &=& i\,\frac{d}{2}\,U^{t}(q)\, \sum_{\mathbf{q}} 
\frac{(q_z^2-q_\perp^2)\,(q_x-i\,q_y)}{q_\perp\,q}
\,F(\mathbf{q},\mathbf{r}),
%\,e^{-i\,{\mathbf{q}}{\mathbf{r}}}\, a_{\mathbf{q}}^\dagger,
\end{eqnarray}
\noindent for transversal $t1$ phonons, and
\begin{eqnarray}
\nonumber
p^{t2} &=& 0, \\
\nonumber
q^{t2} &=& 0, \\
\nonumber
%r^{t2} &=& -i\,U^{t}(q)\, \sum_{\mathbf{q}} 
%\left( \sqrt{3} b\,\frac{q_x\,q_y}{q_\perp} - i\frac{d}{2}\,\frac{q_y^2-q_x^2}{q_\perp} \right) \\
%\nonumber
% & & \,F(\mathbf{q},\mathbf{r}), \\
r^{t2} &=& -i\,U^{t}(q)\, \sum_{\mathbf{q}} 
\left( \sqrt{3} b\,\frac{q_x\,q_y}{q_\perp} - i\frac{d}{2}\,\frac{q_y^2-q_x^2}{q_\perp} \right) \,F(\mathbf{q},\mathbf{r}), \\
%\,e^{-i\,{\mathbf{q}}{\mathbf{r}}}\, a_{\mathbf{q}}^\dagger,\\
s^{t2} &=& -\frac{d}{2}\,U^{t}(q)\, \sum_{\mathbf{q}} 
\frac{q_z\,(q_x - i\, q_y)}{q_\perp}
\,F(\mathbf{q},\mathbf{r}), 
%\,e^{-i\,{\mathbf{q}}{\mathbf{r}}}\, a_{\mathbf{q}}^\dagger,
\label{eq:fin}
\end{eqnarray}
\noindent for transversal $t2$ phonons.

\subsection{Relaxation rate}

We calculate the spin relaxation from an initial hole state $|\Psi_i^h\rangle$,
with energy $E_i^h$, to a final hole state $|\Psi_f^h\rangle$ with energy $E_f^h$.
The relaxation rate is estimated with a Fermi golden rule. 
We consider zero temperature, so that there is no phonon absorption. 
After integrating over phonon degrees of freedom, the rate is given by:
\begin{equation}
\tau_{i \rightarrow f}^{-1} \, = \, 
\frac{2\pi}{\hbar} \, \sum_{\lambda,\mathbf{q}}\,
\left| \langle \Psi_f^h|
{\cal H}_{h-ph}^{\lambda q}\,
| \Psi_i^h \rangle \, \right|^2 \, \delta(\Delta E_{fi}+\hbar c_\lambda q).
%| \Psi_i^h \rangle \, \right|^2 \, \delta(E_f^h-E_i^h+\hbar c_\lambda q).
\label{eq:fgr}
\end{equation}
\noindent Here ${\cal H}_{h-ph}^{\lambda q}$ is the hole-phonon interaction Hamiltonian, 
Eq.~(\ref{eq:h-ph}), but for a fixed phonon branch $\lambda$ and momentum $\mathbf{q}$,
and $\Delta E_{fi}=E_f^h-E_i^h$.
It can be seen from Eqs.~(\ref{eq:ini}-\ref{eq:fin}) that all the terms of ${\cal H}_{h-ph}^{\lambda q}$ 
contain a factor which depends on $\mathbf{q}$ only and $F(\mathbf{q},\mathbf{r})$, 
which depends on spatial coordinates as well.
Thus, when expanded, the matrix element $\langle \Psi_f^h| {\cal H}_{h-ph}^{\lambda q}\, | \Psi_i^h \rangle$
takes the form:
\begin{equation}
\label{eq:G}
\langle \Psi_f^h| {\cal H}_{h-ph}^{\lambda q}\, | \Psi_i^h \rangle = 
 \sum_{J_z',J_z,r',r}\,  (c_{r',J_z'}^f)^*\,c_{r,J_z}^i\,
M_{J_z',J_z}^\lambda(\mathbf{q})\,
\,G_{r',r}(\mathbf{q}).
\end{equation}
\noindent where $G_{r,r'}(\mathbf{q}) = \langle r' |\,e^{-i \mathbf{q} \mathbf{r}}\,| r \rangle$
and $M_{J_z',J_z}^\lambda(\mathbf{q})$ gathers the $\mathbf{q}$-dependent factor of
the ${\cal H}_{h-ph}^{\lambda q}$ term coupling $J_z$ and $J_z'$.
$G_{r,r'}(\mathbf{q})$ is calculated analytically using iterative procedures as described
in Ref.~\cite{PlanellesJPCc}. The sum over $\mathbf{q}$ in Eq.~(\ref{eq:fgr}) is then 
carried out using numerical integration. To this end, it is convenient to use spherical coordinates,
as the modulus $q$ is fixed by the resonance condition and we are left with a two-dimensional 
integral.

Calculations are carried out for InAs QDs embedded in a GaAs matrix. When differences are expected, 
we also calculate GaAs QD embedded in an Al$_x$Ga$_{1-x}$As matrix. Table \ref{tab:par} summarizes 
the parameters we use. The parameters correspond to the QD material, except for the crystal density and 
velocity of sound, which correspond to the matrix material because we assume bulk phonons (for simplicity, 
for Al$_x$Ga$_{1-x}$As we assume $x \rightarrow 0$ and use GaAs phonon parameters). For the dielectric
constant, an average value of $\epsilon_r=12.9$ is taken all over the structure.
The basis used to solve Hamiltonian (\ref{eq:H}) contains all the \emph{hh} oscillator 
eigenstates with the quantum numbers $v_x, v_y < 13$ and $v_z < 9$. 

\begin{table}
\caption{Parameters used in the numerical calculations for InAs (left column) and GaAs (right column) QDs. 
GaAs parameters are used for the matrix in both cases.
$e^-$, $h^+$ and $ph$ stand for electron, hole and phonon.\label{tab:par}} 
\begin{indented}
\item[]
%	\scriptsize
      \begin{tabular}{@{}lllll}
      \br
      Parameter & Symbol & InAs & GaAs & Ref. \\ \mr
      $e^-$ mass ($m_0$) & $m_e$ & 0.026 & 0.067 & \cite{VurgaftmanJAP} \\ 
      $h^+$ Luttinger param. & $\gamma_1$ & 20 & 6.98 & \cite{VurgaftmanJAP} \\ 
      $h^+$ Luttinger param. & $\gamma_2$ & 8.5 & 2.06 & \cite{VurgaftmanJAP} \\ 
      $h^+$ Luttinger param. & $\gamma_3$ & 9.2 & 2.93 & \cite{VurgaftmanJAP} \\ 
      $e^-$ deformation pot. (eV) & $a_{c}$ & -5.08 & -7.17 & \cite{VurgaftmanJAP} \\ 
      $h^+$ deformation pot. (eV) & $a$     &  1.0  &  1.16 & \cite{VurgaftmanJAP} \\ 
      $h^+$ deformation pot. (eV) & $b$     & -1.8  & -2.0 & \cite{VurgaftmanJAP} \\ 
      $h^+$ deformation pot. (eV) & $c$     & -3.6  & -4.8 & \cite{VurgaftmanJAP} \\ 
      $e^-$ BIA coeff. (eV$\cdot$\AA$^3$) & $b_{41}^c$ & 27.18 & 27.58 & \cite{Winkler_book} \\ 
      $h^+$ BIA coeff. (eV$\cdot$\AA) & $C_k$ & -0.0112 & -0.0034 & \cite{Winkler_book} \\ 
      $h^+$ BIA coeff. (eV$\cdot$\AA$^3$) & $b_{41}$ & -50.18 & -81.93 & \cite{Winkler_book} \\ 
      $h^+$ BIA coeff. (eV$\cdot$\AA$^3$) & $b_{42}$ &   1.26 &   1.47 & \cite{Winkler_book} \\ 
      $h^+$ BIA coeff. (eV$\cdot$\AA$^3$) & $b_{51}$ &   0.42 &   0.49 & \cite{Winkler_book} \\ 
      $h^+$ BIA coeff. (eV$\cdot$\AA$^3$) & $b_{52}$ &  -0.84 &  -0.98 & \cite{Winkler_book} \\ 
      Longitudinal $ph$ speed (m/s) & $c_{l}$ & 4720 & 4720 & \cite{VurgaftmanJAP} \\ 
      Transversal $ph$ speed (m/s) & $c_{t}$ & 3340 & 3340 & \cite{VurgaftmanJAP} \\ 
      Crystal density (kg/m$^3$) & $\rho$ & 5310 & 5310 & \cite{VurgaftmanJAP} \\ 
      Piezoelectric coeff. (V/cm) & $h_{14}$ & 3.5$\cdot$10$^6$ & 1.45$\cdot$10$^7$ & \cite{Madelung_book} \\ \br
%      Dielectric constant & $\epsilon_r$ & 15.5 & 12.8 & \cite{Madelung_book} 
      \end{tabular}
%	\normalsize 
\end{indented}
\end{table}

\section{Numerical Results and Discussion}

We shall start this section by describing the dependence of the hole spin lifetime on the
QD geometry and the spin splitting magnitude (Section \ref{ss:geo}).
The influence of each parameter can be understood by analyzing the spin admixture
mechanisms, as we show in Section \ref{ss:mechs}. 
Next, in Section \ref{ss:lh}, we study the effect of the ground state changing from 
mainly \emph{hh} character, which is the case addressed in most previous studies, 
to mainly \emph{lh} character. This transition is observed in QDs with large aspect ratio.\cite{HuSCI,KatzPRL,VoonNL,PlanellesJPCc2}
Last, in Section \ref{ss:mechs_ph}, we compare the role of deformation potential
and piezoelectric potential interactions in determining the efficiency of 
hole-phonon coupling.

\subsection{Geometry and spin splitting dependence}\label{ss:geo}

\begin{figure*}[t]
%\begin{figure}[H]%\label{fig:h_vs_e}
%\begin{figure}[h]
\begin{center}
\includegraphics[width=0.65\textwidth]{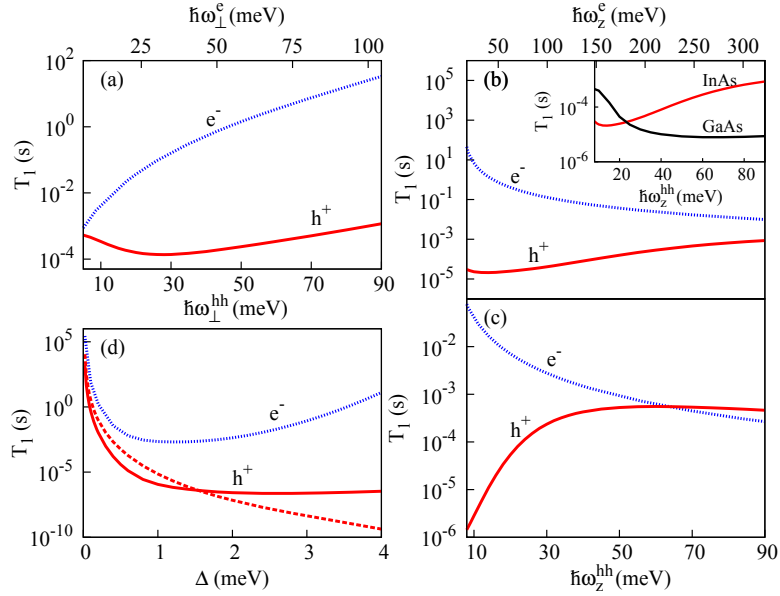} 
\caption{Hole (red solid line) and electron (blue dotted line) spin lifetime in InAs QDs, as a function of 
the lateral confinement (a), vertical confinement (b-c) and spin splitting energy (d).
The inset in (b) compares $T_1^h$ for InAs and GaAs. 
(a): $\hbar \omega_z^{hh}=50$ meV, $\hbar \omega_z^e=179$ meV, $\Delta=0.4$ meV.
(b): $\hbar \omega_\perp^{hh}=20$ meV, $\hbar \omega_\perp^e=23.2$ meV, $\Delta=0.4$ meV.
(c) $\hbar \omega_\perp^{hh}=5$ meV, $\hbar \omega_\perp^e=5.8$ meV, $\Delta=0.4$ meV.
(d): $\hbar \omega^{hh}_\perp=20$ meV and $\hbar \omega^{hh}_z=50$ meV (red solid line);
$\hbar \omega^{e}_\perp=23.2$ meV and $\hbar \omega^{e}_z=179$ meV (blue dotted line);
$\hbar \omega^{hh}_\perp=40$ meV and $\hbar \omega^{hh}_z=5$ meV (red dashed line). % act *** 
}\label{fig:h_vs_e}
\end{center}
\end{figure*}

Solid lines in Figure \ref{fig:h_vs_e} show the hole spin lifetime as a function
of the QD geometry and the spin splitting energy of InAs QDs.
For comparison, we also plot the electron spin lifetimes (dotted lines). The latter have been
calculated using the same formalism as for holes but adapted for single-band conduction electrons.\cite{PlanellesJPCc} 
Both electrons and holes are assumed to be confined in the same QD, hence they share the same force constants
but have different confinement frequencies. % act ***
%Both electrons and holes are assumed to feel the same force constant, but we use
%the \emph{hh} oscillator frequency, $\omega^{hh}_z$ or $\omega^{hh}_\perp$, 
%to denote the confinement strength (for simplicity, we shall often drop the \emph{hh} superscript). 
One can see immediately in the figure that the behavior of holes differs drastically from 
the well-known case of electrons. Below we summarize the influence of each parameter.

Fig.~\ref{fig:h_vs_e}(a) shows the spin lifetime dependence on the lateral confinement in
QDs with strong vertical confinement. % ($\hbar \omega_z^{hh} = 50$ meV).
$T_1^e$ increases monotonically with $\omega_\perp$. This is because, as we approach the spherical
confinement regime ($\omega_z^e=\omega_\perp^e$),  the Dresselhaus SOI of electrons is gradually suppressed.\cite{PlanellesJPCc}
No such trend is however observed for holes, as $H_{BIA}$ does not cancel out even if the 
confinement is isotropic.  % note e.g. in Hb41 that the off-diag ops contain k+ and k- summands, which give different prefactors to kx*ky vs kz^2
As a matter of fact, $T_1^h$ shows an evident non-monotonic behavior, with a minimum at $\omega_\perp^{hh} = 28$ meV and increasing away from it. 
It is worth noting that a previous study by Woods et al. predicted $T_1^h$ to decrease with the QD diameter\cite{WoodsPRB}, 
while a similar study by Lu and co-workers for somewhat larger QDs predicted the opposite trend.\cite{LuPRB} 
Fig.~\ref{fig:h_vs_e}(a) shows that both predictions are conciliable, corresponding to the right and left sides 
of the $T_1^h$ minimum, respectively.
The origin of the different trends will be discussed in Section \ref{ss:mechs}.

%\begin{widetext}
%\end{widetext}

Figs.~\ref{fig:h_vs_e}(b-c) show the spin lifetime dependence on the vertical confinement in QDs 
with moderately strong %($\hbar \omega_\perp^{hh} = 20$ meV, panel (b)) 
and weak %($\hbar \omega_\perp^{hh} = 5$ meV, panel (c)) 
 lateral confinement, respectively. These confinement strengths roughly correspond to
self-assembled (panel (b))\cite{BloklandPRB,WarburtonPRB} and electrostatic (panel (c))\cite{Schinner_arxiv} QDs. % en realidad hw_X0 = 0.8 meV en gated
As can be seen in Fig.~\ref{fig:h_vs_e}(b), electrons and holes have opposite behaviors.
$T_1^e$ now decreases with $\omega_z^e$ because the structure is becoming flatter (less isotropic). 
Instead, the behavior of $T_1^h$ is similar to that observed for varying lateral confinement,
with a shallow minimum at $\omega_z^{hh} = 14$ meV.  
 Previous studies have shown that $T_1^h$ increases with the vertical confinement.\cite{LuPRB}
This is consistent with the right side of the $T_1^h$ minimum in Fig.~\ref{fig:h_vs_e}(b), 
but here we show that the opposite trend is also possible if the QD aspect ratio is large enough 
(left side of the minimum). % el incremento no se ve demasiado bien % tmp ***
%On the other hand, when the lateral confinement is weak, as in Fig.~\ref{fig:h_vs_e}(c), %Fig.~\ref{fig:h_vs_e}(c), 
%the turning points of $T_1^e$ and $T_1^h$ are shifted towards very small $\hbar \omega_z$ values so that only the
%right side behavior is seen. Interestingly, $T_1^h$ now shows a clear saturation with

Fig.~\ref{fig:h_vs_e}(c) illustrates the case in which the lateral confinement is weak.
The minimum of $T_1^h$ is now shifted towards very small $\hbar \omega_z^{hh}$ values so that only the
right side behavior is seen in the range under study. 
Interestingly, here $T_1^h$ shows a clear saturation with increasing vertical confinement ($\hbar \omega_z^{hh} > 40$ meV), 
which has not been noticed before.\cite{LuPRB} As we show in Section \ref{ss:mechs}, this saturation 
reflects the fact that $H_{BIA}$ has replaced \emph{hh-lh} mixing as the main source of SOI.

The inset in Fig.~\ref{fig:h_vs_e}(b) compares $T_1^h$ in InAs and GaAs QDs with
moderate lateral confinement. %($\hbar \omega^{hh}_\perp = 20$ meV) and varying $\hbar \omega^{hh}_z$.
As can be observed, the spin lifetime in InAs QDs is longer than that in GaAs QDs
when $\omega_z^{hh} > \omega_\perp^{hh}$, which is e.g. the case of self-assembled QDs.
This is an unexpected result because in bulk the valence band SOI of InAs is stronger than that of GaAs 
(compare the split-off band splittings\cite{VurgaftmanJAP} or the $\gamma_2$ and $\gamma_3$ coefficients
appearing in $R$ and $S$ terms of Eq.~(\ref{eq:Hl}), which couple \emph{hh} to \emph{lh}).
The underlying reason is that in confined structures, the cubic Dresselhaus SOI interaction becomes important
and it is stronger for GaAs (see $b_{41}$ in Table \ref{tab:par}).
%This is an unexpected result because the valence band SOI of InAs is stronger than that of GaAs.\cite{VurgaftmanJAP}
%The underlying reason is that the Dresselhaus SOI interaction is stronger for GaAs (see $b_{41}$ in Table \ref{tab:par}),
%and it provides a significant contribution to the total spin admixture.  % act ***

Figure \ref{fig:h_vs_e}(d) shows the the spin-flip lifetime of electrons and holes
in a self-assembled-like QD  %($\hbar \omega_z^{hh}=50$ meV, $\hbar \omega_\perp^{hh}=20$ meV) 
as a function of the spin splitting energy. For electrons, $T_1^e$ is largely
determined by the efficiency of the carrier-phonon coupling.\cite{PlanellesJPCc} 
It is short when the phonon wavelength is of the same order as the carrier wavefunction extension,
but it increases for large (small) $\Delta$ values because the phonon wavelength becomes
too short (long), as $q \approx \Delta/\hbar c$. 
 The same happens for holes (notice that $G_{r,r'}(\mathbf{q})$ in Eq.~(\ref{eq:G}) 
vanishes in the limits of large and small phonon wavevector).
Yet, figure \ref{fig:h_vs_e}(d) shows that $T_1^h$ is only sensitive to $\Delta$ 
for small splittings, but then it reaches a plateau where $T_1^h \sim \mu s$. 
The different behavior of holes and electrons is due to the different effective mass
along the growth direction, $m_z$.
For $\hbar \omega_z=50$ meV, the characteristic length of the oscillator states in the
growth direction, $l_z = \sqrt{\hbar / m_z\,\omega_z}$, 
is $l^e_z=4.5$ nm for electrons and $l^{hh}_z=2.4$ nm for holes, i.e. the hole confinement
is stronger. As a result, larger values of $\Delta$ than those used in Fig.~\ref{fig:h_vs_e}(d) % act ***
are required for $T_1^h$ to increase.

Experiments with excitons in self-assembled InGaAs QDs have revealed a negligible 
influence of \emph{e-h} interactions on $T_1^h$.\cite{HallAPL}
Our results would be consistent with such a observation in the regime of large $\Delta$.
For self-assembled QDs, however, $\Delta \leq 0.5$ meV.
Thus, the insensitivity noticed in the experiment is inconsistent with the single-phonon 
processes we consider in Fig.~\ref{fig:h_vs_e}. This suggests that two-phonon processes dominate in these systems.\cite{LiaoPRB,TrifPRL}
On the other hand, experiments with colloidal nanorods have shown a sizable increase of $T_1^h$ 
when changing from type-I to type-II confinement, which modulates the \emph{e-h} exchange
energy from typical colloidal structure values (few meV) to type-II system values (fractions of meV).\cite{HePRL} 
We have run simulations for a nanorod-like geometry (red dashed line in Fig.~\ref{fig:h_vs_e}(d))
and find that the weak vertical confinement renders $T_1^h$ sensitive to $\Delta$ for all 
the range under study, in agreement with the experiment. This indicates that the weak vertical
confinement of rods renders single-phonon processes efficient. 
%Since we obtain qualitatively similar trends to those of Fig.~\ref{fig:h_vs_e}(d) for other geometries
%(including spherical and rod structures), the experiment can be interpreted by assuming that 
%the evolution from type-I system (where $\Delta$ is of few meV) to type-II system (where $\Delta$
%can be fractions of meV) takes $T_1^h$ from the plateau value to the regime of high sensitivity. 

To close this section, we notice %We close this section by remarking that in all cases studied in Fig.~\ref{fig:h_vs_e}, $T_1^e > T_1^h$.
%that previous theoretical studies have predicted that hole spin lifetimes can exceed those of electrons in very flat QDs.\cite{BulaevPRL} 
 that previous theoretical studies with simpler models had predicted that hole spin lifetimes can exceed those of electrons in very flat QDs.\cite{BulaevPRL}  % act ***
%Fig.~\ref{fig:h_vs_e} shows that this could actually be achieved in gated structures, where lateral confinement is very 
 Fig.~\ref{fig:h_vs_e} confirms that this could actually be achieved in gated structures, where lateral confinement is very 
weak (see the crossing between $T_1^e$ and $T_1^h$ in panel (c)).
However, for typical self-assembled InAs QDs,  $T_1^e$ is about one order of magnitude longer
than $T_1^h$ (see panel (b) for large $\hbar \omega_z$).\cite{m=0.045}
This is consistent with experimental measurements by Heiss et al.\cite{HeissPRB}

\subsection{Mechanism of spin admixture}\label{ss:mechs}

Spin admixture is a requirement for spin-flip transitions to take place.\cite{KhaetskiiPRB}
In this section, we compare the spin admixture resulting from all the SOI terms affecting the hole ground state.
As we shall see, once the dominant SOI mechanism is determined, one can rationalize the geometry 
dependence of $T_1^h$ described in the previous section. For convenience of the analysis, in what follows 
we plot and discuss relaxation rates, $1/T_1^h$. Furthermore, we shall often drop the $\emph{hh}$ superscript 
when referring to the confinement frequency, $\omega_\perp^{hh}$ or $\omega_z^{hh}$. % act ***

\begin{figure}[h]
\begin{center}
\includegraphics[width=0.45\textwidth]{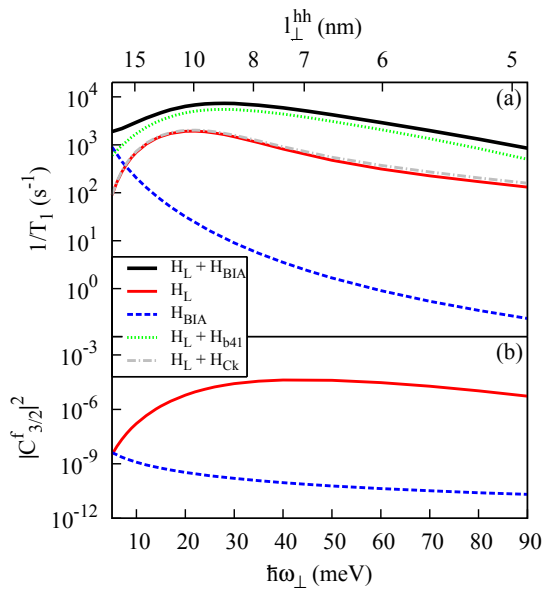}
\caption{Hole spin relaxation (a) and weight of the minor \emph{hh} component (b) as a function of the lateral confinement. 
Different SOI terms are considered. The system is the same as that of Fig.~\ref{fig:h_vs_e}(a).}
\label{fig:mech_wxy}
\end{center}
\end{figure}

Figure \ref{fig:mech_wxy}(a) shows the relaxation rate for the InAs QDs of Fig.~\ref{fig:h_vs_e}(a),
 but now obtained by calculating hole states with the diagonal terms of $H_L$ plus different SOI terms: 
off-diagonal $H_L$ terms (hereafter \emph{hh-lh} coupling), full Dresselhaus Hamiltonian ($H_{BIA}$), 
\emph{hh-lh} coupling plus linear-in-$k$ term ($H_L+H_{C_k}$) and \emph{hh-lh} coupling plus the dominant 
cubic-in-$k$ Dresselhaus term ($H_L+H_{b_{41}}$). 
The total rate, corresponding to $H_L+H_{BIA}$, is also shown (thick black line).
One can see that $H_L$ (red solid line) is more important than $H_{BIA}$ (blue dashed line) 
for large $\omega_\perp$ values. However, as the lateral confinement is weakened and the system becomes 
flatter, $H_{BIA}$ gains weight. % becomes increasingly important.  
For self-assembled QDs ($\hbar \omega_\perp \approx 10-25$ meV), $H_{BIA}$ is already comparable
to $H_L$ and it becomes dominant for weakly confined (e.g. gated) QDs.
Fig.~\ref{fig:mech_wxy} also reveals that the linear-in-$k$ BIA term (gray dot-dashed line)
is but a secondary mechanism, which barely enhances the relaxation rate coming from $H_L$. 
This is inspite of the fact that it is a source of direct admixture between \emph{hh} states with opposite spin, 
with no participation of \emph{lh} states.  For this reason, it had been proposed as the dominant SOI term in 
flat QDs with weak lateral confinement.\cite{TsitsishviliPRB}
Instead, Fig.~\ref{fig:mech_wxy} shows that most of the $H_{BIA}$ contribution comes from the cubic-in-$k$ $b_{41}$ term (see green dotted line). % act ***
 %Instead, most of the $H_{BIA}$ contribution comes from the cubic-in-$k$ $b_{41}$ term (see green dotted line).
This term relies on intermediate \emph{lh} states in order to mix \emph{hh} $J_z=+3/2$ and $J_z=-3/2$ components
(see Eq.~(\ref{eq:Hb41}) in the Appendix), which implies that a simultaneous description of \emph{hh} and \emph{lh} 
states is necessary for realistic modeling. 

By comparing the total relaxation rate coming from $H_L+H_{BIA}$ with that coming from $H_L$ and 
$H_{BIA}$, it is clear that the total rate is not just the sum of the two independent mechanisms.
For example, at $\hbar \omega_\perp=30$ meV, adding $H_{BIA}$ to $H_L$ enhances $1/T_1^h$ about an order 
of magnitude, even though the contribution coming from $H_{BIA}$ alone is about 100 times smaller
than that coming from $H_L$. 
This can be understood by means of a perturbative reasoning: neither $H_L$, Eq.~(\ref{eq:Hl}),
nor $H_{b_{41}}$ -the most relevant term of $H_{BIA}$-, Eq.~(\ref{eq:Hb41}), contribute to 
\emph{hh-lh} mixing at first order. $H_L$ contributes at second order, due to terms involving
non-zero products like $H_L(1,2)\,H_L(2,4)$, while $H_{b_{41}}$ does not. It contributes 
at third order, due to non-zero products like $H_{b_{41}}(1,2)\,H_{b_{41}}(2,3)\,H_{b_{41}}(3,4)$. 
However, the sum of the two Hamiltonians allows $H_{b_{41}}$ to contribute at second order.
Thus, the effect of $H_{b_{41}}$ is clearly non-additive because it is enhanced by $H_L$.
Simultaneously accounting for both SOI terms is then required for quantitative estimates. 

For a qualitative understanding of the geometry dependence of $1/T_1^h$, we rewrite the hole states, 
Eq.~($\ref{eq:psi}$), as $|\Psi^h_m> =  \sum_{J_z} c_{J_z}^m \, |\phi_{J_z}^m\rangle\, |3/2,J_z \rangle$,
where $|\phi_{J_z}^m\rangle=\sum_r\,c_{J_z,r}^m\,|r\rangle$ is the envelope function associated with 
the periodic Bloch function $|3/2,J_z\rangle$.
If we restrict to the diagonal components of $H_{h-ph}$, the matrix element
determining the relaxation rate becomes:

\begin{equation}\label{eq:q1}
\langle \Psi_f^h| {\cal H}_{h-ph}^{\lambda q}\, | \Psi_i^h \rangle 
\approx 
\sum_{J_z}\,(c_{J_z}^f)^* c_{J_z}^i\, \langle \phi_{J_z}^f | {\cal H}_{h-ph}^{\lambda q}\, | \phi_{J_z}^i \rangle.
\end{equation}

\noindent When the QD is oblated or spherical, the low energy states are essentially \emph{hh} states. 
Thus, the initial (final) state of the spin transition is mostly a pure spin up (spin down) \emph{hh} state. 
Considering that $|c_{3/2}^i| \gg  |c_{3/2}^f|$ ($|c_{-3/2}^f| \gg  |c_{-3/2}^i|$), %and Kramers degeneracy, 
%implies $\langle \phi_{J_z}^f | {\cal H}_{h-ph}^{\lambda q}\, | \phi_{J_z}^i \rangle = \langle \phi_{-J_z}^f | {\cal H}_{h-ph}^{\lambda q}\, | \phi_{-J_z}^i \rangle$.
%With these simplifications, one can obtain:
 one can obtain an approximate expression:
\begin{equation}\label{eq:q2}
\frac{1}{T_1^h}  \propto  \left| \langle \Psi_f^h| {\cal H}_{h-ph}^{\lambda q}   | \Psi_i^h \rangle \right|^2 \propto 
  |c_{3/2}^f|^2 \,\left| \langle \phi_{3/2}^f| {\cal H}_{h-ph}^{\lambda q}   | \phi_{3/2}^i  \rangle \right|^2.
\end{equation}
\noindent In other words, the relaxation rate is proportional to the spin admixture of the
ground state through the squared coefficient of the minor \emph{hh} component 
(here spin up, $J_z=3/2$), and proportional to the efficiency of the hole-phonon coupling through 
the rightmost matrix element. %$\langle f^{-3/2}_f| {\cal H}_{h-ph}^{\lambda q}\, | f^{-3/2}_i  \rangle$.

The geometry dependence of $1/T_1^h$ for a given SOI mechanism simply reflects the 
spin admixture variation. This can be seen in Fig.~\ref{fig:mech_wxy}(b), which shows 
that, for $H_L$ and $H_{BIA}$,  $|c_{3/2}^f|^2$ has the same qualitative dependence on
the geometry as the corresponding $1/T_1^h$ values in Fig.~\ref{fig:mech_wxy}(a).
This allows us to interpret the observed maximum as a function of $\omega_\perp$. % act ***
For $H_L$, the spin admixture between the \emph{hh} spin up and spin down components takes 
place necessarily through the intermediate \emph{lh} components, see Eq.~(\ref{eq:Hl}).
The weight of the minor \emph{hh} component is then related to the strength of the off-diagonal 
terms of $H_L$ ($R$ and $S$) and to the energy splitting between the \emph{hh} and the \emph{lh} states.
For small $\omega_\perp$ values, the main effect of increasing the lateral confinement is
to enhance the coupling terms, which are proportional to $k_x$ and $k_y$ (see Eqs.~(\ref{eq:R})-(\ref{eq:S})).
As a result, the weight of the minor \emph{hh} component $|c_{3/2}^f|$ increases, hence $1/T_1^h$ increases.
For larger $\omega_\perp$ values, however, when the lateral and vertical confinements become comparable,
the main effect of increasing the lateral confinement is to bring \emph{lh} states closer to \emph{hh} ones.\cite{VoonNL,PlanellesJPCc2}
When this happens, the \emph{lh} states stop acting as intermediate states for the admixture 
between \emph{hh} components and start participating in the admixture themselves. 
This is at the expense of reducing the weight of the minor \emph{hh} component, hence $1/T_1^h$ decreases.

%The \emph{hh-lh} coupling terms scale up with the confinement strength (note $R$ and 
%$S$ operators dependence on $k$).  
%In turn, the energy splitting, is largely influenced by the different masses of \emph{hh} and \emph{lh}. 
% It is maximum for flat structures and goes to zero for slightly prolate structures.\cite{VoonNL,PlanellesJPCc2} 
%On the left side of Fig.~\ref{fig:mech_wxy}, $\omega_\perp \ll \omega_z$, so \emph{lh} states are 
%far in energy. Changing $\omega_\perp$ has a weak effect on the energy splitting, which is mainly 
%set by the vertical confinement. Then, the main effect of increasing $\omega_\perp$ is to enhance the 
%\emph{hh-lh} coupling strength, resulting in larger $|c_{3/2}^f|$ values. 
%Around $\hbar \omega_\perp \approx 30$ meV, however, the lateral confinement energy becomes comparable 
%to the vertical one. The \emph{lh} states get close in energy and start participating in the admixture 
%themselves, thus reducing the spin admixture of the minor \emph{hh} component.\\

\begin{figure}[h]
\begin{center}
\includegraphics[width=0.45\textwidth]{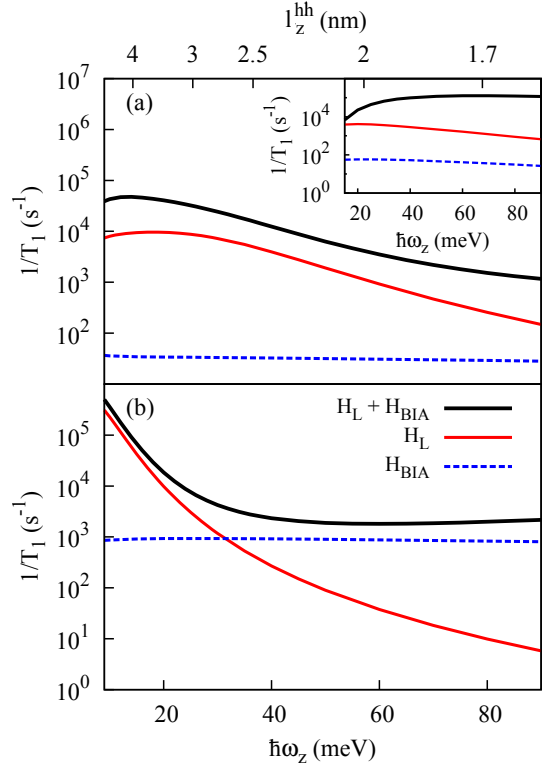}
\caption{Hole spin relaxation as a function of the vertical confinement. Different SOI terms are considered. 
 %Same legend as in Fig.~\ref{fig:mech_wxy} is used. 
(a): strong lateral confinement, $\hbar \omega_\perp=20$ meV. (b): weak lateral confinement, $\hbar \omega_\perp=5$ meV. 
The inset in (a) is the corresponding result for GaAs. The system is the same as that of Figs.~\ref{fig:h_vs_e}(b) and (c).}
\label{fig:mech_wz} 
\end{center}
\end{figure}

Next, we analyze the mechanisms involved in the spin relaxation with varying vertical confinement.
Figure \ref{fig:mech_wz} shows $1/T_1^h$ for the same systems as in Fig.~\ref{fig:h_vs_e}(b-c),
but calculating the hole states with the diagonal terms of $H_L$ plus 
%off-diagonal $H_L$ terms (red solid line) or full $H_{BIA}$ Hamiltonian (blue dashed line).
 \emph{hh-lh} coupling (red solid line) or full $H_{BIA}$ Hamiltonian (blue dashed line).
Panel (a) corresponds to strong lateral confinement. %, $\hbar \omega_\perp=20$ meV (as in Fig.~\ref{fig:h_vs_e}(b)).
The total rate has a maximum at $\hbar\omega_z = 14$ meV, 
whose origin is analogous to that described above for varying lateral confinement.
Beyond the maximum, the total rate ($H_L + H_{BIA}$) decreases with the vertical confinement strength,
in agreement with previous studies.\cite{LuPRB} Yet, we also find that the decrease eventually saturates.
This is evident for InAs QDs with weak lateral confinement, as shown in Fig.~\ref{fig:mech_wz}(b),
or GaAs QDs with strong lateral confinement, as in Fig.~\ref{fig:mech_wz}(a) inset. 
The origin of this saturation is the contribution of $H_{BIA}$, which provides a
lower bound to $1/T_1^h$. In particular, $H_{b_{41}}$ introduces off-diagonal coupling 
terms which are quadratic in $k_z$ (see operator $L_{41}$ in the Appendix),
instead of the linear $k_z$ terms of $H_L$ (see $S$ operator in Eq.~(\ref{eq:Hl})). 
Since the uncoupled \emph{hh} and \emph{lh} energies are also quadratic in $k_z$,
a perturbational analysis easily shows that the two contributions compensate each other.
For strong $\omega_z$, when lateral confinement is negligible, the cancellation is exact and
the relaxation rate does not depend on the vertical confinement.

%\noindent where $\eta_1'$ and $\eta_2'$ are mass dependent constants. One can see that
%for $\omega_z \gg \omega_\perp$, the $\omega_z$ factors in the numerator and denominator cancel out
%and the rate does not depend on the vertical confinement. This is because the off-diagonal terms of
%$H_{b_{41}}$ are quadratic in $k_z$, which is the same dependence as that of the energy splitting.

\begin{figure}[h]
\begin{center}
\includegraphics[width=0.45\textwidth]{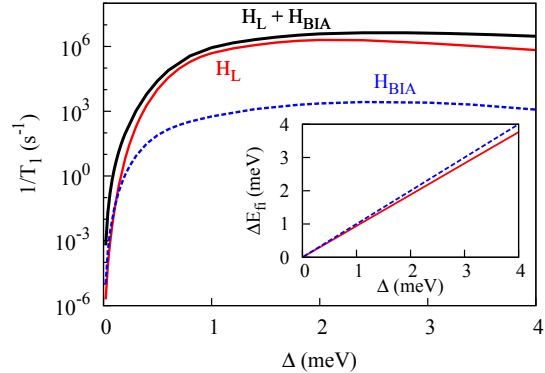}
\caption{Hole spin relaxation as a function of the spin splitting energy in a QD with $\hbar \omega_\perp=20$ meV and $\hbar \omega_z=50$ meV. 
Same legend as in Fig.~\ref{fig:mech_wxy} is used.
 %Thick black line: $H_L+H_{BIA}$, red solid line: $H_L$, blue dashed line: $H_{BIA}$. 
The inset compares the energy splitting between the Kramer doublet for \emph{hh-lh} coupling and Dresselhaus SOI.} 
\label{fig:mech_split} 
\end{center}
\end{figure}

The magnitude of the spin splitting also influences the dominant mechanism of spin admixture.
This is illustrated in Fig.~\ref{fig:mech_split}, which considers a self-assembled InAs QD
with varying spin splitting energy.  For large $\Delta$, $H_L$ has a dominant contribution 
to the relaxation rate, but $H_{BIA}$ becomes equally important for small enough $\Delta$. 
The relative enhancement of the role of $H_{BIA}$ originates in its zero-field spin splitting,
 which leads to larger energy difference between the Kramers pair ($\Delta E_{fi}$) 
than with $\Delta$ alone, as shown in the figure inset. When $\Delta \rightarrow 0$
and the phonon wavelength increases beyond the QD size, the extra energy coming from the zero-field 
spin splitting of $H_{BIA}$ provides a significant contribution to preserve the hole-phonon coupling efficiency.

\subsection{Light hole spin relaxation}\label{ss:lh}

\begin{figure}[h]
\begin{center}
\includegraphics[width=0.45\textwidth]{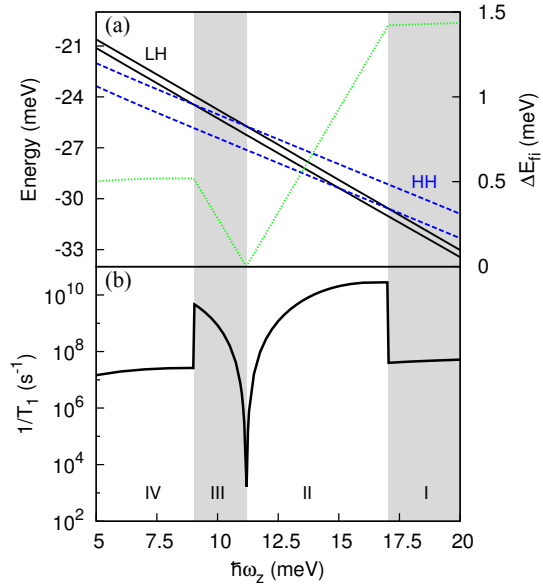} 
\caption{Hole energy levels (a) and spin relaxation rate (b) around the \emph{hh-lh} crossing region.
$\hbar \omega_\perp=40$ meV and $\Delta=2$ meV. In (a), solid and dashed lines are used for states
with dominant \emph{lh} and \emph{hh} character, respectively. The dotted line gives the energy splitting.
Shades are used to identify the regions with different kind of states involved in the transition.}\label{fig:lh}
\end{center}
\end{figure}

When the aspect ratio increases and the QD shape becomes prolate, the hole ground state evolves from the eminent \emph{hh}
character discussed so far to \emph{lh} character, as noted e.g. in colloidal nanorods.\cite{HuSCI,KatzPRL,VoonNL,PlanellesJPCc2}
Here we investigate how the change of the ground state affects the relaxation between the
two highest hole states. Fig.~\ref{fig:lh}(a) shows the hole energy levels in a QD
with $\hbar \omega_\perp=40$ meV as a function of $\hbar \omega_z$.
In the limit of strong and weak vertical confinement, the two highest states are essentially
\emph{hh} and \emph{lh} doublets, $J_z=\pm 3/2$ and $J_z=\pm 1/2$ respectively.
In the intermediate regime, $\hbar \omega_z = 9-17$ meV, the two doublets cross.
This gives rise to pronounced changes in the relaxation rate, as shown in Fig.~\ref{fig:lh}(b).

The changes can be understood as follows. In region I of the figure, the two highest states
are the \emph{hh} doublet. The relaxation is between states with opposite spin and roughly
constant energy splitting (see $\Delta E_{fi}$, dotted line in Fig.~\ref{fig:lh}(a)).
At $\hbar \omega_z=17$ meV, when we enter region II, the excited \emph{hh} state crosses with 
the first \emph{lh} state. Now the relaxation is between a \emph{lh} ($J_z=-1/2$) and a \emph{hh} ($J_z=-3/2$).
%Since \emph{lh} have mixed spin up and spin down projections, there is no need for spin flip and the 
%transition is much faster. 
Since \emph{lh} have mixed spin up and spin down projections, there is no need for spin flip.
Then, the $s^\lambda$ terms of the strain Hamiltonian, $H_{dp}$, provide direct coupling with \emph{hh} 
and the resulting transition is much faster.
This explains the abrupt increase of $1/T_1^h$. However, the energy splitting between
the \emph{hh} and \emph{lh} becomes smaller with decreasing $\omega_z$ because 
of their different masses. As a result, at $\hbar \omega_z=11$ meV, the \emph{lh} replaces
the \emph{hh} as the ground state. Near the degeneracy point, $\hbar \omega_z=11.2$ meV, 
$\Delta E_{fi}$ is so small that hole-phonon coupling becomes inefficient and the relaxation is strongly suppressed, 
but it increases again in region III for the same reasons as in region II. 
Finally, at $\hbar \omega_z=9$ meV, the excited \emph{lh} state crosses with the highest
\emph{hh} state. We thus enter region IV, where the transition takes place between two \emph{lh} states with 
orthogonal Bloch functions, $|3/2,\pm 1/2\rangle$. $J_z$ admixture mechanisms are necessary and the 
relaxation becomes slow. As a matter of fact, the spin relaxation timescale for transitions between 
\emph{lh} states is similar to that between \emph{hh} states, in spite of the fact that their Bloch 
functions contain mixed spins.

\subsection{Mechanism of hole-phonon coupling}\label{ss:mechs_ph}

Electron-acoustic phonon coupling in QDs is known to occur mainly through deformation potential 
interaction when the energy splitting is large ($\Delta E_{fi} > 0.1$ meV) and through piezoelectric 
potential when it is small.\cite{ClimentePRB} 
In principle, for holes the situation may differ because the deformation potential Hamiltonian, 
Eq.~(\ref{eq:dp}), is formally different from that of electrons.  To investigate this point, in this section 
we compare the role of the two kinds of carrier-phonon coupling mechanism for holes subject to varying 
effective magnetic fields.

\begin{figure}[h]
\begin{center}
\includegraphics[width=0.45\textwidth]{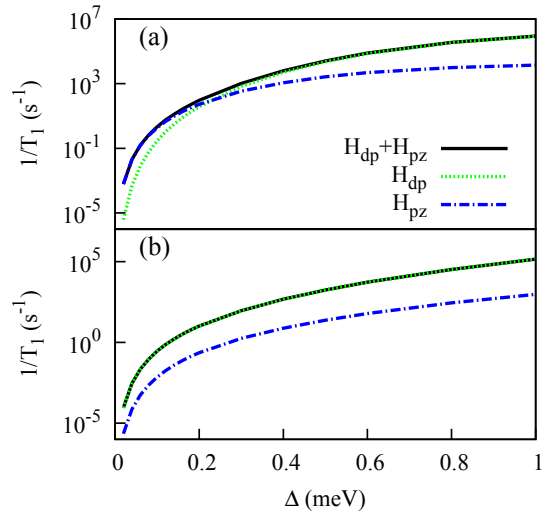} 
\caption{Hole spin relaxation as a function of the spin splitting energy.
(a): transition between \emph{hh} states in a QD with $\hbar \omega_\perp=20$ and $\hbar \omega_z=50$ meV.
(b): transition between \emph{lh} states in a QD with $\hbar \omega_\perp=40$ and $\hbar \omega_z=5$ meV.
}\label{fig:h-ph} 
\end{center}
\end{figure}

Figure \ref{fig:h-ph}(a) shows the spin relaxation rate in an oblate (quasi-2D) QD, 
where the highest states are \emph{hh}, while Fig.~\ref{fig:h-ph}(b) shows that in
a prolate (quasi-1D) QD, where the highest states are \emph{lh}.
For the spin transition within the \emph{hh} doublet, panel (a), the behavior is analogous to
that of electrons. Deformation potential interaction (dotted line) provides the largest contribution to $1/T_1^h$
except for very small $\Delta$, when piezoelectric interaction (dashed-dotted line) takes over.
This is because all the terms of ${\cal H}_{dp}$ are proportional to the phonon momentum $\sqrt{q}$, see 
Eqs.~(\ref{eq:dpl}-\ref{eq:fin}), while the piezoelectric potential is proportional to $1/\sqrt{q}$,
see Eq.~(\ref{eq:ini}).
With decreasing $\Delta$ both mechanisms become inefficient, because for long phonon wavelength 
$e^{i\mathbf{q} \mathbf{r}} \approx 1$. Then, in Eq.~(\ref{eq:G}), the matrix element $G_{r,r'}(\mathbf{q}) \approx \langle r'| r\rangle$,
i.e. it tends to the overlap between the envelope components of the initial and final states.
For \emph{hh}, these components have different symmetries, so the coupling vanishes. For example, 
in axially symmetric structures, the $J_z=+3/2$ component of the initial state has a well defined 
azimuthal angular momentum $m_z=0$, which couples through the $s^\lambda$ operator of $H_{dp}$ 
with the $J_z=+1/2$ component of the final state, for which $m_z=-2$.\cite{mz}

The situation for \emph{lh} is quite different. As shown in Fig.~\ref{fig:h-ph}(b), now
deformation potential interaction is the dominant relaxation channel even for small $\Delta$.
The underlying reason is that, in contrast to the \emph{hh} case, the off-diagonal terms of 
${\cal H}_{dp}$ couple envelope components with the same symmetry. For example, the 
$J_z=+3/2$ and $J_z=+1/2$ components of
the initial and final state now have both $m_z=-1$, and hence are not orthogonal. As a consequence, 
$G_{r,r'}(\mathbf{q})$ does not vanish when $q \rightarrow 0$.

\section{Summary}

We have investigated hole spin relaxation in InAs and GaAs QDs using 3D 4-band k$\cdot$p
Hamiltonians.
We have shown that the hole spin lifetime has a non-monotonic dependence on the lateral 
and vertical confinement strength. This is due to the interplay between the energy splitting
of \emph{hh} and \emph{lh}, which is set by their different masses, and the anisotropic
\emph{hh-lh} coupling terms. The resulting geometry dependence of hole spin relaxation is 
qualitatively different from that of electrons. 

\emph{hh-lh} coupling and Dresselhaus SOI have been found to have comparable contributions
to the spin admixture of hole states in self-assembled QDs, with the former becoming dominant for 
prolate structures, such as nanorods, and the latter for strongly oblate ones, such as gated
QDs. The cubic-in-$k$ Dresselhaus term leads to an upper bound of $T_1^h$ with increasing vertical 
confinement, which is missed when considering \emph{hh}-\emph{lh} coupling only.

We have also investigated the spin relaxation involving states with dominant \emph{lh} character.
Transitions between \emph{lh} and \emph{hh} states are very fast because \emph{lh} have
strong spin admixture. Instead, transitions between \emph{lh} states with different Bloch
angular momentum $J_z$ are as slow as the transitions between \emph{hh} states with opposite spin. 
There is however a difference in the dominating hole-phonon interaction mechanism. 
At small energy splittings, the relaxation is mainly due to deformation potential interaction, 
unlike for \emph{hh} transitions, where it is due to piezoelectric interaction.

\ack
Support from MICINN project CTQ2011-27324,
UJI-Bancaixa project P1-1B2011-01, the Ramon y Cajal program (JIC) 
and UJI fellowship (CS) is acknowledged.\\

\appendix

\section{Dresselhaus Hamiltonian for Holes}

In this appendix we write the explicit matrix forms of the different $H_{BIA}$ terms in cartesian coordinates.
Separating the different invariants in Eq.~(\ref{eq:Hbia}) we obtain:
\begin{equation}
H_{BIA} = H_{C_k} + H_{b_{41}} + H_{b_{42}} + H_{b_{51}} + H_{b_{52}},
\end{equation}
\noindent where:
\begin{equation}\label{eq:HCk}
H_{C_k}\,=\,C_k\,
\left( 
\begin{array}{cccc} 
0                         &  -\frac{k_-}{2}          &   k_z                      &  -\frac{\sqrt{3}\,k_-}{2} \\
-\frac{k_+}{2}            &   0                      &   \frac{\sqrt{3}\,k_+}{2}  &   -k_z  \\
k_z                       &  \frac{\sqrt{3}\,k_-}{2} &   0                        &  -\frac{k_-}{2}   \\
-\frac{\sqrt{3}\,k_+}{2}  & -k_z                     &  -\frac{k_+}{2}            & 0
\end{array}
\right),
\end{equation}
\noindent with $k_\pm = k_x \pm i\,k_y$.
\begin{equation}\label{eq:Hb41}
H_{b_{41}} \,=\,b_{41}\,
\left( 
\begin{array}{cccc} 
\frac{3}{2}\,P_{41}  &  \frac{\sqrt{3}}{2}\,L_{41} &   0                                    &  0 \\
\frac{\sqrt{3}}{2}\,L^{\dagger}_{41}     &  \frac{1}{2}\,P_{41}        &   L_{41}                               &  0 \\
0                    &   L^{\dagger}_{41}          & -\frac{1}{2}\,P_{41}                   &  \frac{\sqrt{3}}{2}\,L_{41} \\
0                    &   0                         &  \frac{\sqrt{3}}{2}\,L^{\dagger}_{41}  & -\frac{3}{2}\,P_{41} 
\end{array}
\right),
\end{equation}
\noindent where $P_{41} = (k_x^2-k_y^2)\,k_z$ and $L_{41} = i\,k_-\,k_x\,k_y - k_+\,k_z^2$.
\begin{equation}\label{eq:Hb42}
H_{b_{42}} \,=\,b_{42}\,
\left( 
\begin{array}{cccc} 
\frac{27}{8}\,P_{41}  &  \frac{7\sqrt{3}}{8}\,L_{41} &   0                                    &  -\frac{3}{4}\,L_{42} \\
\frac{7\sqrt{3}}{8}\,L^{\dagger}_{41}     &  \frac{1}{8}\,P_{41}        &   \frac{5}{2}\,L_{41}                               &  0 \\
0                    &   \frac{5}{2}\,L^{\dagger}_{41}          & -\frac{1}{8}\,P_{41}                   &  \frac{7\sqrt{3}}{8}\,L_{41} \\
-\frac{3}{4}\,L_{42}^\dagger  &   0                         &  \frac{7\sqrt{3}}{8}\,L^{\dagger}_{41}  & -\frac{27}{8}\,P_{41} 
\end{array}
\right),
\end{equation}
\noindent where $L_{42} = i\,k_+\,k_x\,k_y + k_-\,k_z^2$.
\begin{equation}\label{eq:Hb51}
H_{b_{51}} \,=\,b_{51}\,
\left( 
\begin{array}{cccc} 
%0                    & -\frac{\sqrt{3}}{4}\,K_+     &   \frac{\sqrt{3}}{2}\,K_z                   &  -\frac{3}{4}\,K_- \\
%-\frac{\sqrt{3}}{4}\,K_+^\dagger  &  0     &   \frac{3}{4}\,K_+  &  -\frac{\sqrt{3}}{2}\,K_z                 \\
%\frac{\sqrt{3}}{2}\,K_z^\dagger       &   \frac{3}{4}\,K_+^\dagger &  0        &  -\frac{\sqrt{3}}{4}\,K_+  \\
%-\frac{3}{4}\,K_-^\dagger  &  -\frac{\sqrt{3}}{2}\,K_z^\dagger       &  -\frac{\sqrt{3}}{4}\,K_+^\dagger  & 0
0                    & -\frac{\sqrt{3}}{4}\,K_+     &   \frac{\sqrt{3}}{2}\,K_z                   &  -\frac{3}{4}\,K_- \\
-\frac{\sqrt{3}}{4}\,K_-  &  0     &   \frac{3}{4}\,K_+  &  -\frac{\sqrt{3}}{2}\,K_z                 \\
\frac{\sqrt{3}}{2}\,K_z       &   \frac{3}{4}\,K_- &  0        &  -\frac{\sqrt{3}}{4}\,K_+  \\
-\frac{3}{4}\,K_+  &  -\frac{\sqrt{3}}{2}\,K_z       &  -\frac{\sqrt{3}}{4}\,K_-  & 0
\end{array}
\right),
\end{equation}
\noindent where $K_+=K_x+i\,K_y$, $K_-=K_x-i\,K_y$, $K_x = k_x\,(k_y^2+k_z^2)$,  
$K_y = k_y\,(k_x^2+k_z^2)$, and $K_z = k_z\,(k_x^2+k_y^2)$.
\begin{equation}\label{eq:Hb52}
H_{b_{52}} \,=\,b_{52}\,
\left( 
\begin{array}{cccc} 
%0                    & -\frac{\sqrt{3}}{4}\,M_+     &   \frac{\sqrt{3}}{2}\,k_z^3                   &  -\frac{3}{4}\,M_- \\
%-\frac{\sqrt{3}}{4}\,M_+^\dagger  &  0     &   \frac{3}{4}\,M_+  &  -\frac{\sqrt{3}}{2}\,k_z^3                 \\
%\frac{\sqrt{3}}{2}\,(k_z^3)\dagger       &   \frac{3}{4}\,M_+^\dagger &  0        &  -\frac{\sqrt{3}}{4}\,M_+  \\
%-\frac{3}{4}\,M_-^\dagger  &  -\frac{\sqrt{3}}{2}\,(k_z^3)^\dagger       &  -\frac{\sqrt{3}}{4}\,M_+^\dagger  & 0
0                    & -\frac{\sqrt{3}}{4}\,M_+     &   \frac{\sqrt{3}}{2}\,k_z^3                   &  -\frac{3}{4}\,M_- \\
-\frac{\sqrt{3}}{4}\,M_-  &  0     &   \frac{3}{4}\,M_+  &  -\frac{\sqrt{3}}{2}\,k_z^3                 \\
\frac{\sqrt{3}}{2}\,k_z^3       &   \frac{3}{4}\,M_- &  0        &  -\frac{\sqrt{3}}{4}\,M_+  \\
-\frac{3}{4}\,M_+  &  -\frac{\sqrt{3}}{2}\,k_z^3       &  -\frac{\sqrt{3}}{4}\,M_-  & 0
\end{array}
\right),
\end{equation}
\noindent where $M_+=k_x^3+i\,k_y^3$ and $M_-=k_x^3-i\,k_y^3$.


\begin{thebibliography}{53}
\bibitem{m=0.045}
The same conclusions, albeit with somewhat lower $T_1^e$, are obtained using larger
electron masses to account e.g. for Ga diffusion into the InAs QD.
\bibitem{mz} In axially symmetric systems, the envelope functions of the Luttinger 
spinor have well defined z-component of the orbital angular momentum, $m_z=F_z-J_z$.
In oblate QDs, the highest doublet has $F_z=\pm 3/2$, and in prolate QDs it has $F_z=\pm 1/2$.
See e.g. Ref.~\cite{PlanellesJPCc2}.
\end{thebibliography}
\end{document}